\documentclass[reprint,prl,twocolumn,nofootinbib,aps,nobalancelastpage]{revtex4-2}
\usepackage{amsmath}
\usepackage{amsfonts}
\usepackage{amssymb}
\usepackage{mathtools}
\usepackage{mathrsfs}
\usepackage{graphicx}
\usepackage{xcolor}
\usepackage[colorlinks=True,citecolor=gray,linkcolor=black,urlcolor=gray]{hyperref}
\usepackage{hypcap}
\usepackage{braket}
\usepackage{yfonts}
\usepackage{soul}
\usepackage{bm}
\newcommand\eq[1]{\begin{align}#1\end{align}}

\definecolor{myBlue}{RGB}{31,119,180}
\definecolor{myOrange}{RGB}{255,127,14}
\definecolor{myGreen}{RGB}{44,160,44}
\definecolor{myRed}{RGB}{214,39,40}
\definecolor{myPurple}{RGB}{148,103,189}

\makeatletter
\def\p@figure{\color{myBlue}}
\def\p@equation{\color{myRed}}
\makeatother

\begin{document}
\title{Eigenvector Correlations Across the Localisation Transition in non-Hermitian Power-Law Banded Random Matrices}

\author{Soumi Ghosh}
\email{soumi.ghosh@icts.res.in}
\affiliation{International Centre for Theoretical Sciences, Tata Institute of Fundamental Research, Bengaluru 560089, India}

\author{Manas Kulkarni}
\email{manas.kulkarni@icts.res.in}
\affiliation{International Centre for Theoretical Sciences, Tata Institute of Fundamental Research, Bengaluru 560089, India}

\author{Sthitadhi Roy}
\email{sthitadhi.roy@icts.res.in}
\affiliation{International Centre for Theoretical Sciences, Tata Institute of Fundamental Research, Bengaluru 560089, India}

\begin{abstract}
The dynamics of non-Hermitian quantum systems have taken on an increasing relevance in light of quantum devices which are not perfectly isolated from their environment. The interest in them also stems from their fundamental differences from their Hermitian counterparts, particularly with regard to their spectral and eigenvector correlations. These correlations form the fundamental building block for understanding the dynamics of quantum systems as all other correlations can be reconstructed from it. In this work, we study such correlations across a localisation transition in non-Hermitian quantum systems. As a concrete setting, we consider non-Hermitian power-law banded random matrices which have emerged as a promising platform for studying localisation in disordered, non-Hermitian systems. We show that eigenvector correlations show marked differences between the delocalised and localised phases. In the delocalised phase, the eigenvectors are strongly correlated as evinced by divergent correlations in the limit of vanishingly small complex eigenvalue spacings. On the contrary, in the localised phase, the correlations are independent of the eigenvalue spacings. We explain our results in the delocalised phase by appealing to the Ginibre random matrix ensemble. On the other hand, in the localised phase, an analytical treatment sheds light on the suppressed correlations, relative to the delocalised phase. Given that eigenvector correlations are fundamental ingredients towards understanding real- and imaginary-time dynamics with non-Hermitian generators, our results open a new avenue for characterising dynamical phases in non-Hermitian quantum many-body systems.
\end{abstract}

\maketitle

Ergodicity or lack thereof, manifested in localisation, in disordered, interacting quantum many-body systems is a question of immanent interest~\cite{QChaos_DAlessio.2016,deutsch2018eigenstate,review_Nandkishore.2015,review_Alet.2018,review_Abanin.2019,abanin2017recent}. As many-body localised (MBL) systems fail to thermalise under their dynamics, they raise fundamental questions with regard to the statistical mechanical description as well as the precise nature of their dynamics when thrown out of equilibrium (see Refs.~\cite{review_Nandkishore.2015,review_Alet.2018,review_Abanin.2019,review_Abanin.2019,abanin2017recent} for reviews on MBL and further references therein). Conventionally, these questions have been studied in the context of closed quantum systems where the dynamics is unitary.

More recently however, understanding the dynamics of interacting quantum many-body systems described by non-Hermitian Hamiltonians has emerged as an extremely relevant question~\cite{metz2019spectral,NHMBL_hamazaki.2019,Panda_NHMBL.2020,NHMBL-QP_Zhai.2020,Ghosh2022spectral,hamazaki2022lindbladian,ashida2020non,nakagawa2021exact,hamazaki2020universality,yamamoto2022universal}. This is, in part, due to the advent of NISQ devices~\cite{preskill2018quantum,bharti2021noisy,ippoliti2021manybody,lau2022nisq}, wherein the non-Hermiticity induced by external noise, or coupling to environments or measurement apparatuses, is inevitable and understanding its effect is of utmost importance. From a theoretical point of view, the interest lies in their fundamental differences from their Hermitian counterparts owing to the former's complex eigenvalue spectrum. This offers the possibility of realising phase structures of quantum systems quite different than those in Hermitian systems~\cite{Grobe.1988,ashida2020non,ProseLS.2019,luitz2019exceptional,cspacing_lucas.2020,Lucas_Lindblad.2020,wang2020hierarchy,Xiao2022Levelstatistics,csr_syk2022,PhysRevA.105.L050201,PhysRevLett.126.166801,PhysRevB.101.014204,PhysRevB.102.064212,mcginley2022absolutely,garcia2022dominance,garcia2022replica,jia2022replica,detomasi2022nonhermitian}.

\begin{figure}[!t]
    \centering
    \includegraphics[width=\linewidth]{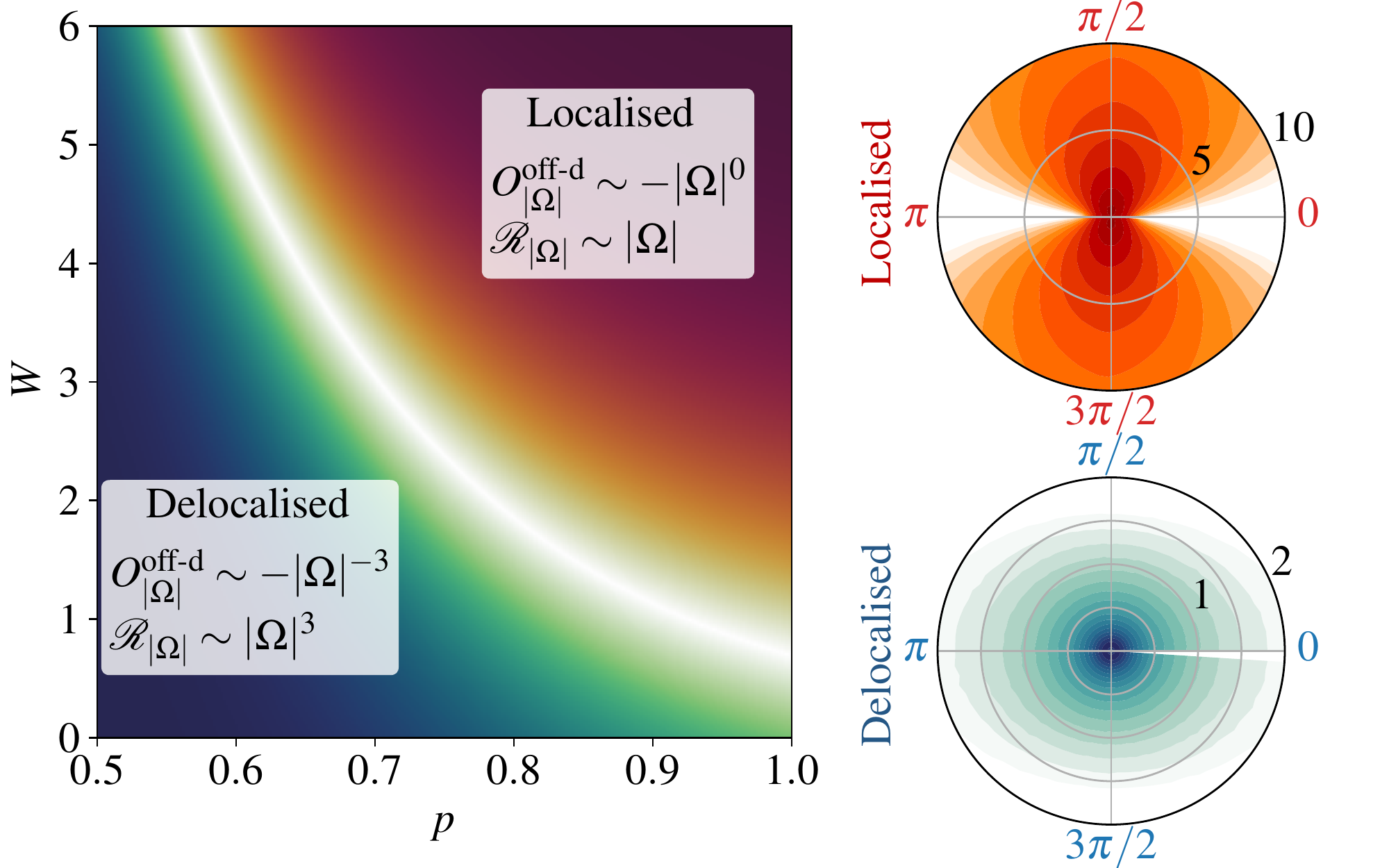}
    \caption{(Left) Schematic phase diagram of non-Hermitian power-law banded random matrix (NH-PLBRM) ensemble \textcolor{black}{in the $p$-$W$ plane, where $W$ denotes the disorder strength of the {\it complex} diagonal elements and $p$ is the exponent of the power-law decay of off-diagonal, hopping matrix elements.} The behaviours of the eigenvector and spectral correlations in the two phases are summarised. (Right) The eigenvector correlations (Eq.~\ref{eq:Odef}) as a heatmap in the plane of complex eigenvalue spacings, in the localised (top) and delocalised (bottom) phases. Besides the difference in their scaling with $|\Omega|$, another stark difference is that in the delocalised phase the correlations are isotropic whereas in the localised phase, we find that they are strongly anisotropic.
    } 
    \label{fig:schematic}
\end{figure}

Under the umbrella of dynamics of non-Hermitian quantum systems, the physics of non-Hermitian many-body localisation and the associated localisation transition has been under intense investigation of late~\cite{NHMBL_hamazaki.2019,Panda_NHMBL.2020,NHMBL-QP_Zhai.2020,Ghosh2022spectral,yamamoto2023localization}.Spectral and eigenvector correlations constitute the basic building block for a theory of dynamics of any quantum system as all other dynamical correlations can be reconstructed from it.
As far as spectral properties are considered, the ergodic phase of such systems displays universality, manifested in level repulsion~\cite{NHMBL_hamazaki.2019,NHMBL-QP_Zhai.2020} in the complex eigenvalue spectrum as well as a ramp in the dissipative spectral form factor~\cite{dsff_Li.2021,Ghosh2022spectral,Shivam2023manybody}, akin to Ginibre random matrix ensembles~\cite{ginibre1965statistical,byun2023progressI,byun2023progressII}. On the other hand, the spectral properties in the localised phase show starkly different behaviour and deviate significantly from random matrix behaviour. In fact, these as well as the participation ratios of eigenstates~\cite{Zeng2017Anderson,Schiffer2021Anderson,Suthar2022nonHermitian}, which are a measure of how (de)localised the eigenstates are, have been extremely insightful diagnostics of the ergodicity or localisation in disordered, non-Hermitian systems.
However, one of the most fundamental ingredients to get a complete understanding of the dynamics of quantum systems are dynamical eigenvector correlations. While they have been studied extensively for Hermitian systems across localisation transitions~\cite{fyodorov1997strong,tikhonov2019statistics,tikhonov2021eigenstate}, they have been hitherto unexplored in non-Hermitian settings with results available only for random matrices~\cite{chalker1998eigenvector,mehlig2000statistical}. This leads us to the central motivation of our work, namely, the behaviour of eigenvector correlations across localisation transitions in non-Hermitian systems.

As a concrete setting, we use power-law banded random matrices (PLBRM), but in their non-Hermitian incarnation. For Hermitian systems, PLBRMs have long been used as an archetypal model for localisation transitions in quantum systems~\cite{levitov1990delocalisation,levitov1989absence,mirlin1996transition,mirlin2000multifractality,evers2000fluctuations,evers2008anderson}. In a very recent work, non-Hermitian power-law banded random matrices (NH-PLBRM) were also shown to exhibit localisation transitions~\cite{detomasi2023nonhermiticity}. In fact, NH-PLBRMs were shown to exhibit localisation in parameter regimes where localisation is forbidden in their Hermitian counterparts.

Our results show that eigenvector correlations (as well as spectral correlations) show stark differences in the delocalised and localised phases, see Fig.~\ref{fig:schematic} for a summary of our main findings. In the delocalised phase, we find that the results for the correlations fall in the universality class of Ginibre random matrices~\cite{chalker1998eigenvector,mehlig2000statistical,Crawford2022Eigenvector}. An appropriately defined correlation between eigenvectors diverges as the complex eigenvalue spacing decreases, suggesting that the eigenvectors are very strongly correlated. 
By contrast, in the localised phase, we find that the correlations are independent of the spacing between the eigenvalues at small spacings which suggests that the correlations are strongly suppressed relative to those in the delocalised phase. We explain this behaviour via an analytical calculation based on a simple perturbation theory where the bare resonances are renormalised appropriately. Within the limits of our numerical calculations we find an anomalous, intermediate behaviour of the correlations in the critical regime. 

The importance of our results lies in that the transient dynamics of non-Hermitian systems are controlled by the eigenvector correlations. Our results constitute a firm step towards understanding the spectral and dynamical properties of local observables across localisation transitions in disordered, interacting, non-Hermitian quantum many-body systems.

To set the stage formally, consider a $N\times N$ non-Hermitian Hamiltonian matrix, $H$, with complex eigenvalues, $z_\alpha$. The corresponding left and right eigenvectors, $\bra{L_{\alpha}}$ and $\ket{R_{\alpha}}$, which satisfy
\eq{
    \bra{L_\alpha}H = \bra{L_\alpha}z_\alpha;\quad\quad H\ket{R_\alpha} = z_\alpha\ket{R_\alpha}\,,
}
form a complete, biorthonormal set with $\braket{L_\alpha|R_\beta}=\delta_{\alpha\beta}$. Requiring that eigenvector correlations are invariant under scale transformations, the simplest non-trivial measure of the correlations can be defined as~\cite{chalker1998eigenvector} 
\eq{
    O_{\alpha\beta} = \braket{L_\alpha|L_\beta}\braket{R_\beta|R_\alpha}\,.
    \label{eq:Oalphabeta}
}
The definition in Eq.~\ref{eq:Oalphabeta} directly implies that $O_{\alpha\beta}=O_{\beta\alpha}^\ast$, and also from completeness, $\sum_{\alpha}O_{\alpha\beta}=1$.
It will be useful to resolve the correlations in Eq.~\ref{eq:Oalphabeta} in terms of the eigenvalues, and define averaged diagonal and offdiagonal correlations, $O^\mathrm{d}$ and $O^\text{off-d}$, respectively as 
\eq{
    O^\mathrm{d}(z) =& \Braket{N^{-1}\sum_\alpha O_{\alpha\alpha}\delta(z-z_\alpha)}\,,\label{eq:Od-def}\\
    O^\text{off-d}(Z,\Omega) =& \bigg\langle N^{-1}\sum_{\alpha\neq\beta} O_{\alpha\beta}\delta\left(Z-\frac{z_\alpha+z_\beta}{2}\right)\times\nonumber\\&\quad\quad\quad\quad\quad\quad\quad\quad\delta(\Omega-z_\alpha+z_\beta)\bigg\rangle\,.
    \label{eq:Odef}
}
The offdiagonal correlation, as defined above, depends on both the mean of the eigenvalues, $Z$, as well as their difference $\Omega$. However, for simplicity, we will be interested in two specific versions of it. The first is where the mean is integrated over,
\eq{
    O^\text{off-d}(\Omega)&\equiv \int dZ~O^\text{off-d}(Z,\Omega) \nonumber\\ &= \Braket{N^{-1}\sum_{\alpha\neq\beta} O_{\alpha\beta}\delta(\Omega-z_\alpha+z_\beta)}\,,
    \label{eq:O-Omega}
}
and the second is where we restrict the sum over pairs of eigenvectors in Eq.~\ref{eq:Odef} such that the mean of their eigenvalues is vanishing,
\eq{
    O^\text{off-d}_{Z=0}(\Omega)\equiv O^\text{off-d}(Z=0,\Omega)\,.\label{eq:O-Z0-Omega}
}
Note that the averaged eigenvector correlations in Eqs.~\ref{eq:O-Omega} and \ref{eq:O-Z0-Omega} are functions of $\Omega$ which is complex. In much of the following, we will find that it is sufficient to consider and focus on the respective correlations as a function of $|\Omega|$. With $\Omega=|\Omega|e^{i\theta}$, they are defined as
\eq{
    \tilde{O}^\text{off-d}_{|\Omega|}(|\Omega|) = |\Omega|\int_0^{2\pi} d\theta~O^\text{off-d}(\Omega)\,,
    \label{eq:O-mod-Omega}
}
and similarly for $\tilde{O}^\text{off-d}_{Z=0,|\Omega|}$.

As we will show later, both $\tilde{O}^\text{off-d}_{|\Omega|}$ as well as $\tilde{O}^\text{off-d}_{Z=0,|\Omega|}$ exhibit the same universal behaviour at small $|\Omega|$. However, this universal behaviour is starkly different between delocalised and localised phases. In particular, in the thermodynamic limit $N\to\infty$ for $|\Omega|\ll 1$, in the delocalised phase both $\tilde{O}^\mathrm{offd}_{|\Omega|}, \tilde{O}^\mathrm{offd}_{Z=0,|\Omega|}\sim -|\Omega|^{-3}$ whereas in the localised phase, we find that both of them scale $\sim -|\Omega|^0$. This constitutes the central result of this work.

It is important to note here that if $H$ were to be Hermitian, the eigenvector correlation defined in Eq.~\ref{eq:Oalphabeta} would have been trivial with $O_{\alpha\beta}=\delta_{\alpha\beta}$. As such the diagonal correlation $O^\mathrm{d}$ in Eq.~\ref{eq:Od-def} would have simply been the density of states and the off-diagonal ones in Eq.~\ref{eq:Odef} would have been identically zero. The non-triviality in the correlations arises purely from the non-hermiticity. However, the crucial point is that the nature of the correlations depends on the phase in which the non-Hermitian system lies.

\begin{figure}[!t]
    \includegraphics[width=0.475\textwidth]{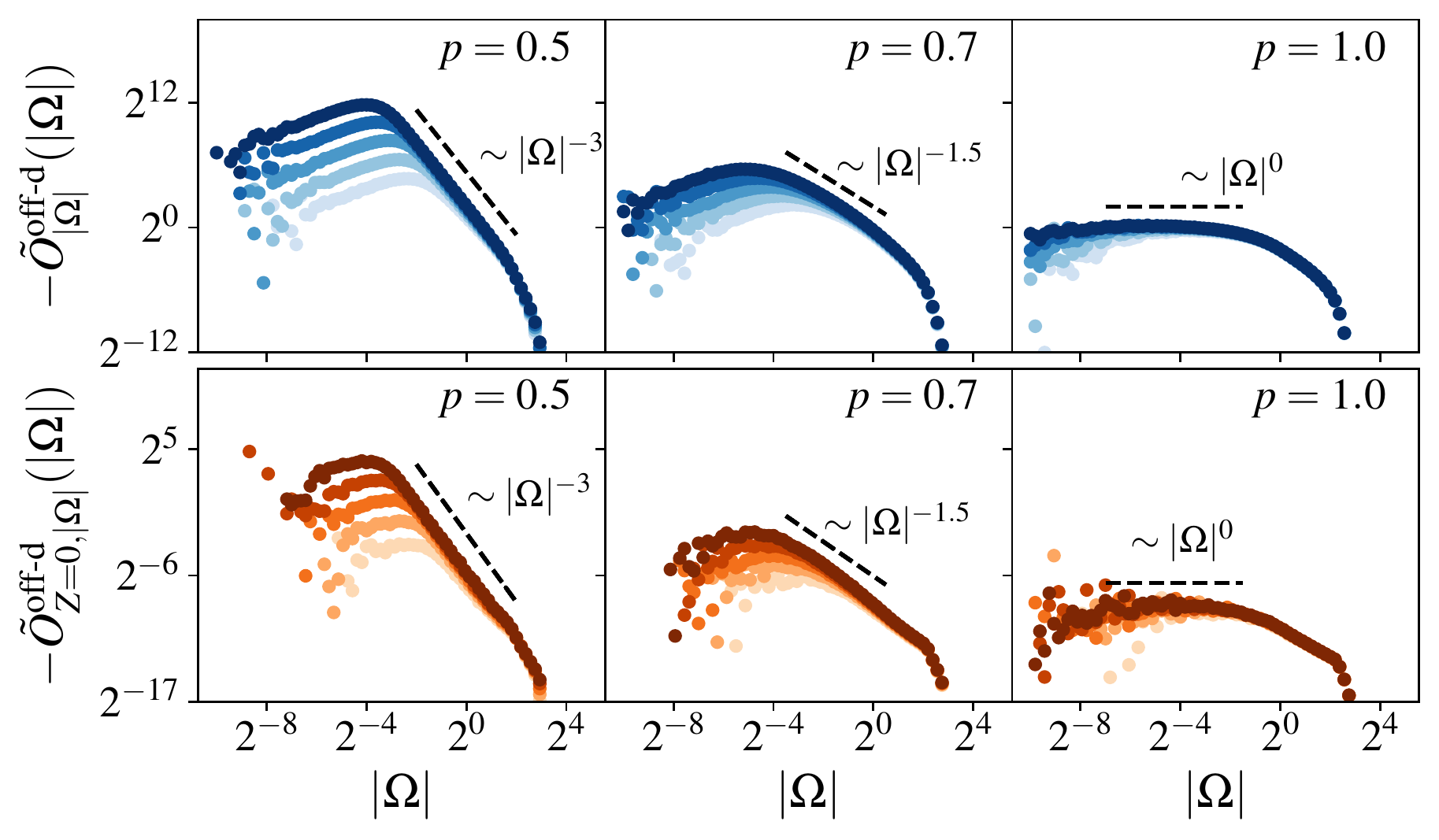}
    \caption{Off-diagonal eigenvector correlations for three different values of $p$ in the delocalised (left), critical (centre), and localised (right) regime. The top and bottom rows correspond to $\tilde{O}^{\text{off-d}}_{|\Omega|}$ (Eq.~\ref{eq:O-mod-Omega}) and $\tilde{O}^{\text{off-d}}_{Z=0,|\Omega|}$ respectively. The dashed lines indicate the power laws, $|\Omega|^{-3}$ (delocalised), $|\Omega|^{-1.5}$ (critical), and $|\Omega|^0$ (localised). The different colour intensities correspond to different system sizes, $N=128,256,512,1024, 2048$ (lighter to darker). Data is for $W=3$.} \label{fig:offdiagonal}
\end{figure}

We now delve into the details of our results and start with describing the NH-PLBRM ensemble~\cite{detomasi2023nonhermiticity}. In an instance of the Hamiltonian from the ensemble, the element $H_{mn}$ is given by
\eq{
H_{mn} = \epsilon_n \delta_{mn}+j_{mn}\,,
\label{eq:Hamiltonian}
}
where $j_{nm}^*=j_{mn}$ and $\epsilon_n$, $j_{mn}$ are complex random numbers. The real and imaginary parts of the diagonal elements are both chosen from uniform distributions; $\mathrm{Re}[\epsilon_n],\mathrm{Im}[\epsilon_n]\in [-W,W]$.
The real ($\mathrm{Re}$) and imaginary ($\mathrm{Im}$) parts of the independent off-diagonal elements $j_{mn}~(m>n)$  are chosen from uniform distributions,
$\mathrm{Re}[j_{mn}],\mathrm{Im}[j_{mn}]\in [-\sigma_{|m-n|},\sigma_{|m-n|}]$. The width $\sigma_r$ decays with $r=|m-n|$ following a power law,
\eq{ \sigma_r=[2(r^2+b^2)]^{-p/2}\,,
\label{eq:sigr}
}
where $b$ is the bandwidth of the decay ($b\simeq 1$) and $p$ is the power of the off-diagonal power-law decay term.
Here the imaginary parts of the diagonal elements bring about the  non-Hermiticity in the Hamiltonian. In the absence of those, the random matrices become Hermitian and the chaotic limit of the underlying model corresponds to the Gaussian unitary ensemble (strictly speaking, complex Wigner matrix ensemble since they are sampled from uniform distributions). In an analogous way, the chaotic limit of the NH-PLBRM is shown to correspond to the Ginibre unitary ensemble (GinUE)~\cite{ginibre1965statistical}. 
The NH-PLBRM has a rich localisation phase diagram in the $p$-$W$ plane (see Fig.~\ref{fig:schematic}). 
{\color {black} While the model was introduced and studied in detail in Ref.~\cite{detomasi2023nonhermiticity}, we summarise its salient features for completeness. Unlike its Hermitian counterpart where localisation is forbidden for $p<1$~\cite{evers2008anderson}, the NH-PLBRM hosts a localised phase and a disorder driven localisation transition for $1/2\leq p\leq 1$. However, the localised phase is algebraic in nature, again in contrast to the Hermitian PLBRM. Finally, we note that the NH-PLBRM does not host a localised phase for $p<1/2$ which can be understood via simple resonance counting argument~\cite{detomasi2023nonhermiticity}~\footnote{\color{black} Since the complex energies live on a two-dimensional plane, the mean-level spacing of sites at distance $r$ from any given site $\sim W/r^{1/2}$ which when compared to the power-law decaying matrix element ($\sim 1/r^p$) implies that localisation is forbidden for $p<1/2$.}.}
Also, we find that the density of states of the NH-PLBRM in the complex eigenvalue plane is uniform (wherever finite) to a very good approximation~\cite{supp} which lets us conveniently avoid spectrum unfolding while defining the eigenvector correlations in Eq.~\ref{eq:Odef}.

The numerical results for the eigenvector correlations, obtained from exact diagonalisation (ED) of the Hamiltonians in Eq.~\ref{eq:Hamiltonian}, are shown in Fig.~\ref{fig:offdiagonal}. The top row corresponds to $\tilde{O}^\text{off-d}_{|\Omega|}$ defined in Eq.~\ref{eq:O-mod-Omega} and the bottom row to $\tilde{O}^\text{off-d}_{Z=0,|\Omega|}$. In the delocalised phase (left column), we find that both of them scale as $-|\Omega|^{-3}$. By contrast, in the localised phase (right column) they scale approximately as $-|\Omega|^{0}$. In the critical regime between the two phases ($p\approx 0.7$ for $W=3$), our numerical results suggest an anomalous scaling $\sim -|\Omega|^{-1.5}$.
\begin{figure}
    \includegraphics[width=0.475\textwidth]{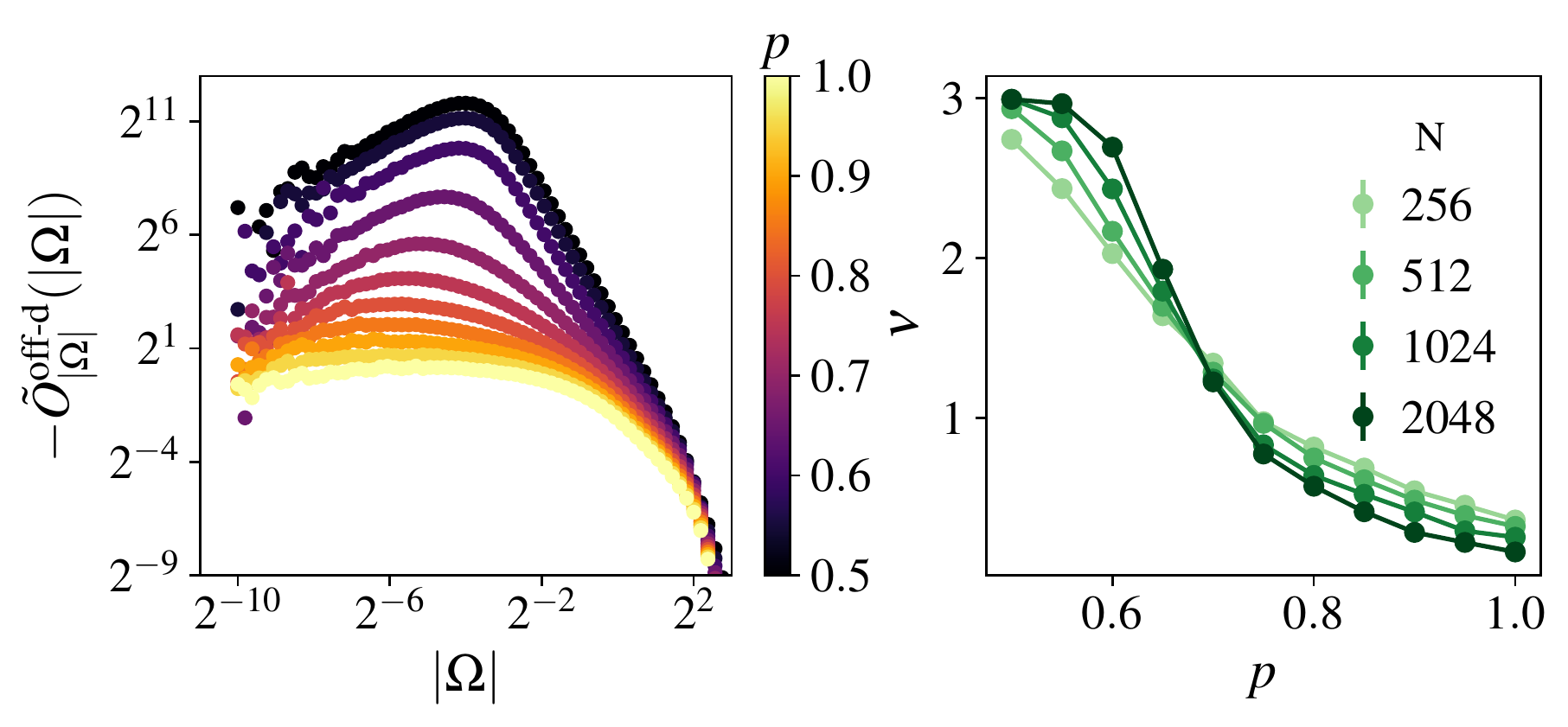}
    \caption{(Left) Off-diagonal correlations as a function of $|\Omega|$ for different values of $p$ for $N=2048$ and $W=3$. As we move from the delocalised to the localised phase by increasing $p$ from $0.5$ to $1$ (as indicated by the colour-bar), the correlation gets suppressed and the exponent $\nu$ (defined via $\tilde{O}^\text{off-d}_{|\Omega|}\sim -|\Omega|^{-\nu}$) changes from $3$ to $0$.  (Right) Variation of the exponent $\nu$ as a function of $p$ for different $N$ showing a crossing at the putative critical point $p_c\approx 0.7$ with $\nu_c\approx 1.5$.}
    \label{fig:off-diagonal-expoenets}
\end{figure}
\begin{figure}[!b]
    \includegraphics[width=0.475\textwidth]{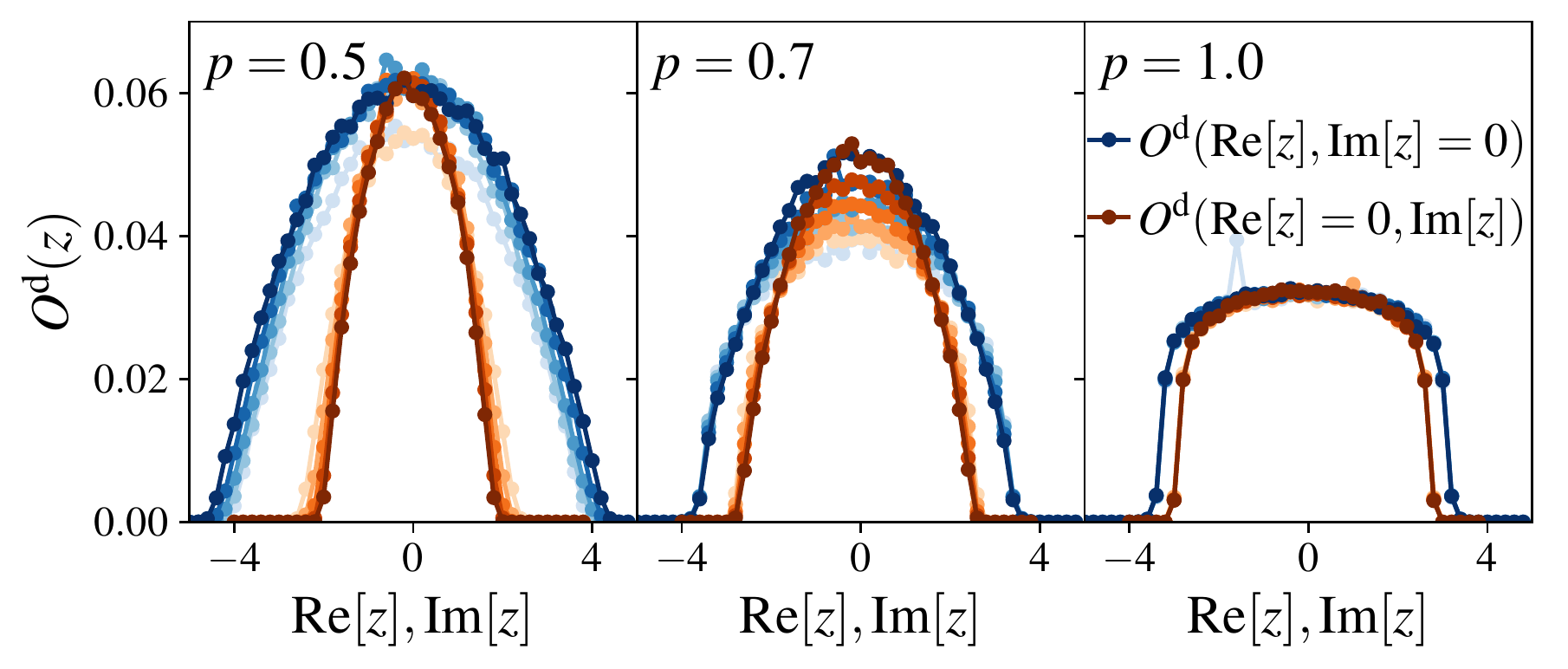}
    \caption{Diagonal correlations $O^{\text{d}}(z)$ defined in Eq.~\ref{eq:Od-def} in the delocalised (left), critical (centre), and the localised (right) regime. The blue data points show the variation of $O^{\text{d}}(\mathrm{Re}[z],\mathrm{Im}[z]=0)$ with $\mathrm{Re}[z]$, while the orange lines show the variation of $O^{\text{d}}(\mathrm{Re}[z]=0,\mathrm{Im}[z])$ with $\mathrm{Im}[z]$. As in Fig.~\ref{fig:offdiagonal}, the different colour intensities indicate different system sizes $N$.}
    \label{fig:diagonal}
\end{figure}
In Fig.~\ref{fig:off-diagonal-expoenets} (left), we show the off-diagonal correlation $\tilde{O}^\text{off-d}_{|\Omega|}$ for a fixed $N=2048$ and $W=3$ but for several values of $p \in [0.5,1.0]$ straddling the critical point at $p_c\approx 0.7$. For a finite system, we observe that $\tilde{O}^\text{off-d}_{|\Omega|}\sim -|\Omega|^{-\nu}$ where the exponent $\nu$ sharply changes from $\nu = 3$ in the delocalised phase to vanishingly small ($\nu \to 0$) in the localised phase. The right panel shows the variation of $\nu$ with $p$ for several values of $N$ with the data for different $N$ showing a crossing at the putative critical point.

While our main focus is on the off-diagonal eigenvector correlations, we also find sharp distinctions in the diagonal correlations, $O^\text{d}(z)$ (defined in Eq.~\ref{eq:Od-def}) between the two phases, as shown in Fig.~\ref{fig:diagonal}. In the delocalised phase (left column) we find an inverted parabolic profile of $O^\text{d}$ symptomatic of GinUE universality~\cite{chalker1998eigenvector}. By contrast, in the localised phase (right), $O^\text{d}$ is significantly flatter and approximately mirrors the density of states profile. This can be understood as deep inside the localised phase the eigenvectors are sharply localised around $\mathcal{O}(1)$ nearby sites which gives rise $O_{\alpha\alpha}\sim \mathcal{O}(1)$ for all $\alpha$ irrespective of its eigenvalue. In the critical regime (centre), we again observe an intermediate behaviour in similar spirit as the off-diagonal correlations.

Having established the numerical results for the eigenvector correlations, we next provide analytical insights into the results for the off-diagonal correlations in both, the delocalised and localised phases. The delocalised phase of the NH-PLBRM can be understood by appealing to the GinUE universality class. The off-diagonal eigenvector correlations in GinUE matrices are given by~\cite{chalker1998eigenvector,mehlig2000statistical}
\eq{
O_{\text{GinUE}}^\text{off-d}(Z,\Omega)=&\frac{Z_+ Z_-^\ast-1}{\pi^2|\Omega|^4}\Theta\left(1-\left|Z_+\right|\right)\Theta\left(1-\left|Z_-\right|\right)\,,
\label{eq:OZOmega-GinUE}
}
where $Z_\pm = Z \pm \Omega/2$. In the limit of $|\Omega|\ll 1$, Eq.~\ref{eq:OZOmega-GinUE} can be used to obtain
\eq{
\tilde{O}^\text{off-d}_{|\Omega|},\tilde{O}^\text{off-d}_{Z=0,|\Omega|} \sim -|\Omega|^{-3}\,,
}
which explains our results in the delocalised phase of the NH-PLBRM and demonstrates that it indeed lies in the GinUE universality class.

\begin{figure}[!b]
    \includegraphics[width=0.475\textwidth]{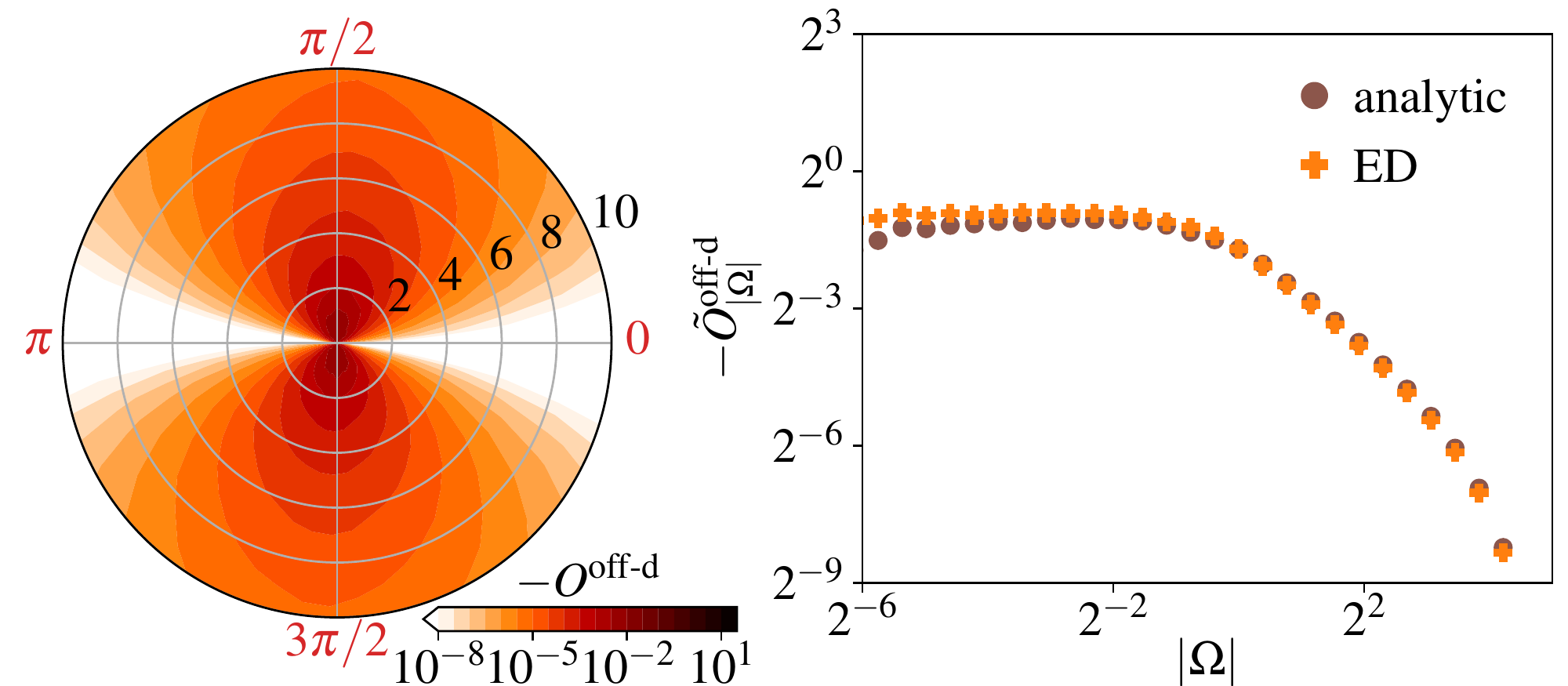}
    \caption{(Left) $O^\text{off-d}(\Omega)$ in the localised phase obtained from the analytic theory (from Eq.~\ref{eq:Yr} and Eq.~\ref{eq:ROmega}) as a colour-map in the complex $\Omega$ plane. (Right) Comparison of $\tilde{O}^\text{off-d}_{|\Omega|}$ as a function of $|\Omega|$ from the analytic calculation with that from ED. Results are for deep in the localised phase with $W=10$, $\alpha=1$, and $N=1024$.}
    \label{fig:loc-analytics}
\end{figure}

Deep in the localised phase, the eigenvectors can be well approximated by leading order perturbative corrections to the site-localised states at infinite disorder. Denoting by $\ket{\alpha}$ a state localised on a single site~\footnote{Deep in the localised phase, since every eigenvector is expected to be closely tied to a site, we use the same notation to index the sites and eigenvectors}, the eigenvectors to leading order are given by
\eq{\label{eq:RLpert}
\ket{R_\alpha} = \ket{\alpha}+\sum_{\gamma\neq\alpha}\frac{H_{\alpha\gamma}}{\Delta_{\alpha\gamma}}\ket{\gamma},\,
\bra{L_\alpha} = \bra{\alpha}+\sum_{\gamma\neq\alpha}\frac{H_{\gamma\alpha}}{\Delta_{\alpha\gamma}}\bra{\gamma},
}
where $\Delta_{\alpha\gamma}=\epsilon_\alpha-\epsilon_\gamma$. Also, at leading order, $z_\alpha=\epsilon_\alpha$. Using  Eq.~\ref{eq:RLpert} and the definition in Eq.~\ref{eq:Oalphabeta}, we obtain
\eq{\label{eq:Opert}
    O^{\mathrm{loc}}_{\alpha\beta} &= -4|H_{\alpha\beta}|^2 (\mathrm{Im}[\Delta_{\alpha\beta}^{-1}])^2\,,
}
for the eigenvector overlaps in the localised phase. Since the expression in Eq.~\ref{eq:Opert} is obtained from an unrenormalised perturbative expansion (Eq.~\ref{eq:RLpert}), \textcolor{black}{it allows for bare resonances due to $\mathrm{Im}[\Delta_{\alpha\beta}^{-1}]\to \infty$} which can result in a divergent overlap. While a mathematically rigorous renormalised perturbation theory, for example {\it \`a la} Feenberg~\cite{feenberg1948note}, is outside the scope of this work, we account for the bare resonances by imposing an empirical cutoff on the overlaps. Physically, this corresponds to setting the $O_{\alpha\beta}$ for the resonant pairs to an empirical $\mathcal{O}(1)$ threshold which is what a proper renormalisation of the resonances would have done self-consistently, and leave the other $O_{\alpha\beta}$'s as they are. To this end, we define a renormalised overlap as 
\eq{\label{eq:Galphabeta}    
G_{\alpha\beta} = O^\mathrm{loc}_{\alpha\beta}\Theta(1-|O^\mathrm{loc}_{\alpha\beta}|)-\Theta(|O^\mathrm{loc}_{\alpha\beta}|-1)\,,
}
and compute the off-diagonal correlations as $O^\text{off-d}(\Omega)=\braket{N^{-1}\sum_{\alpha\neq\beta} G_{\alpha\beta}\delta(\Omega-\Delta_{\alpha\beta})}$, and similarly for $O^\text{off-d}_{Z=0}$. Since the matrix elements of the Hamiltonian, $\{H_{\alpha\beta}\}$ and $\{\epsilon_\alpha\}$, are independent of each other, the eigenvector correlation can be expressed as $O^\text{off-d}(\Omega) = \sum_{r=1}^{N-1} Y_r(\Omega)$ where
\eq{
Y_r(\Omega) = \int d\epsilon_\alpha &P_\epsilon(\epsilon_\alpha)\int d\epsilon_\beta P_\epsilon(\epsilon_\beta)\bigg[\delta(\Omega-\Delta_{\alpha\beta})\times\nonumber\\&\int d H_r P_{H_r}(H_r) \tilde{G}(H_r,\Delta_{\alpha\beta})\bigg]\,.
\label{eq:Yrdef}
}
with $\tilde{G}(H_r,\Delta_{\alpha\beta}) \equiv G_{\alpha\beta}$ and $H_{\alpha\beta}$ set to  $H_r$. 
The notation $H_r$ refers to an hopping matrix element of the Hamiltonian between sites separated by distance $r$ such that it is random complex number with real and imaginary parts drawn from uniform distributions, $\mathrm{Re}[H_r],\mathrm{Im}[H_r]\in [-\sigma_r,\sigma_r]$ where $\sigma_r$ is given by Eq.~\ref{eq:sigr}. Using the distributions for the $\epsilon_
\alpha$'s and $H_r$'s, we obtain~\footnote{For simplicity of expressions, we used a circular distribution of $\epsilon_\alpha$'s and $H_{\alpha\beta}$'s. However, using numerical integration of Eq.~\ref{eq:Yrdef}, we checked that the results in the universal $|\Omega|\ll 1$ regime are identical for a box distribution.} 
\eq{
\label{eq:Yr}
Y_r(\Omega) = \begin{cases}
R(\Omega)\left(1-\frac{1}{8 (\sigma_r\mathrm{Im}[\Omega^{-1}])^2}\right); \sigma_r\ge (2|\mathrm{Im}[\Omega^{-1}]|)^{-1},\\
2 R(\Omega)\sigma_r^2(\mathrm{Im}[\Omega^{-1}])^2;\sigma_r< (2|\mathrm{Im}[\Omega^{-1}]|)^{-1}\,,
\end{cases}
}
where $R(\Omega)$ is the probability that two uncorrelated random $\epsilon$'s are separated by $\Omega$ and it is given by 
\eq{
\label{eq:ROmega}
R(\Omega)=\frac{2}{\pi^2 W^2}\left[\mathrm{cos}^{-1}\left(\frac{|\Omega|}{2W}\right)-\frac{|\Omega|}{2W}\sqrt{1-\left(\frac{|\Omega|}{2 W}\right)^2}\right] \,,
}
with $P_\epsilon(\epsilon) = (\pi W^2)^{-1}\Theta(W-|\epsilon|)$.
Using Eq.~\ref{eq:Yr}, an analytical expression for $O^\text{off-d}(\Omega)$ in the localised phase can be obtained. It, however, is rather cumbersome and opaque and hence we omit it for brevity. Instead, we plot the result for $O^\text{off-d}(\Omega)$ obtained analytically using Eq.~\ref{eq:Yr} and Eq.~\ref{eq:ROmega} as a colour-map in the complex $\Omega$ plane in Fig.~\ref{fig:loc-analytics} (left). The qualitative features of the exact ED result (see Fig.~\ref{fig:schematic}) are well captured. For a quantitative comparison, we derive the corresponding $\tilde{O}^\text{off-d}_{|\Omega|}(|\Omega|)$ and plot it in Fig.~\ref{fig:loc-analytics} (right); we find a remarkable agreement with the exact numerical results and the analytic calculation does indeed yield the approximately $|\Omega|$-independent behaviour of $\tilde{O}^\text{off-d}_{|\Omega|}(|\Omega|)$ at small $|\Omega|$ . 

To conclude, we demonstrated that eigenvector correlations are starkly different between delocalised and localised phases in disordered, non-Hermitian systems. Using NH-PLBRM as a prototype, we showed, via extensive numerical calculations and analytical arguments, that eigenvectors are strongly correlated in the delocalised phase and the same are suppressed in the localised phase (see Fig.~\ref{fig:schematic} for a summary). While eigenvector correlations were the focus of this work, for the sake of completeness, we also calculated spectral correlations, which were characterised by the presence and absence of complex level repulsion in the delocalised and localised phase respectively~\cite{supp}. 

Our findings will have a significant bearing on the characterisation of dynamical phases of non-Hermitian, locally interacting quantum many-body Hamiltonians~\cite{eigvecorr-nh-spin}. While eigenvector correlations, such as the ones discussed here, are definitely interesting in this context, it is equally interesting to understand the spectral properties of local observables in the same spirit. In particular, these quantities are expected to play a pivotal role in understanding fundamental issues like (i) eigenstate thermalisation  (or lack thereof) in non-Hermitian systems~\cite{cipolloni2023entanglement,cipolloni2023nonhermitian} and (ii) non-Hermitian many-body localisation.
In fact, extending these ideas to open quantum systems in general, such as via the eigenvector correlations of the underlying Liouvillian operators~\cite{cspacing_lucas.2020,wang2020hierarchy,denisov2019universal,lange2020zufallsmatrixtheorie,Lucas_Lindblad.2020,lange2021random,garcia2022,sa2022symmetry,orgad2022dynamical}, is topically interesting.

\begin{acknowledgements}
M.K. would like to acknowledge support from the project 6004-1 of the Indo-French Centre for the Promotion of Advanced Research (IFCPAR) and SERB Matrics Grant (MTR/2019/001101) from the Science and Engineering Research Board (SERB), Department of Science and Technology (DST), Government of India. M.K. and S.R. acknowledge support of the Department of Atomic Energy, Government of India, under Project No. 19P1112R\&D. S.R. also acknowledges support from an ICTS-Simons Early Career Faculty Fellowship via a grant from the Simons Foundation (677895, R.G.).
\end{acknowledgements}

\bibliography{refs,ref-prb}

\clearpage

\onecolumngrid
\begin{center}
\textbf{Supplementary Material: Eigenvector Correlations Across the Localisation Transition in non-Hermitian
Power-Law Banded Random Matrices}\\
\medskip

Soumi Ghosh, Manas Kulkarni, and Sthitadhi Roy\\
{\it International Centre for Theoretical Sciences, Tata Institute of Fundamental Research, Bengaluru 560089, India}
\end{center}

\setcounter{equation}{0}
\setcounter{figure}{0}
\setcounter{page}{1}
\renewcommand{\theequation}{S\arabic{equation}}
\renewcommand{\thefigure}{S\arabic{figure}}
\renewcommand{\thesection}{S\arabic{section}}
\renewcommand{\thepage}{S\arabic{page}}

\twocolumngrid

In this supplementary material, we present results for the spectral (complex eigenvalues) properties both in the delocalised and localised phases of non-Hermitian power-law banded random matrix (NH-PLBRM) ensemble.

\subsection*{Density of States}
\begin{figure}[!b]
\includegraphics[width=0.5\textwidth]{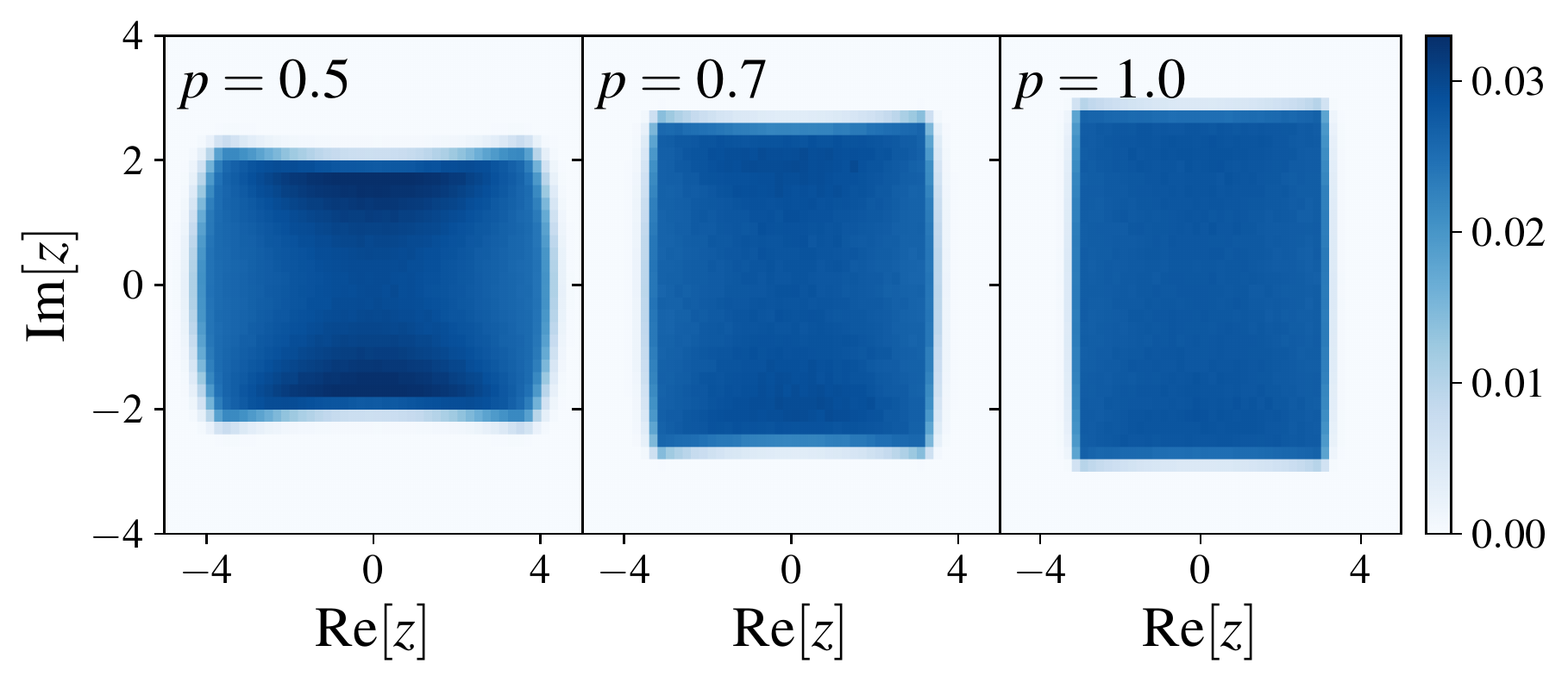}
\centering
\caption{Density of states, $\rho(z)$, in the complex eigenvalue plane as colour-maps for NH-PLBRMs. The three panels correspond to parameters in the delocalised phase (left), critical regime (centre), and the localised phase (right) respectively. The support of $\rho(z)$ is independent of $N$. Data shown here is for $N=1024$.}
\label{fig:dos}
\end{figure}

The normalised density of states in the complex eigenvalue plane is defined as usual 
\eq{
\rho(z) = \frac{1}{N}\Braket{\sum_{\alpha=1}^N\delta(z-z_\alpha)}\,.
\label{eq:rhoz}
}
We show representative results for $\rho(z)$ in the three regimes, delocalised, critical, and localised, in Fig.~\ref{fig:dos}. The central message is that $\rho(z)$ is uniform to a very good approximation over the values of $z$ on which it is supported. As mentioned, in the main text, it conveniently lets us avoid non-trivial spectrum unfolding while defining eigenvector correlations.

\subsection*{Spectral correlations}

\begin{figure}
\includegraphics[width=0.4\textwidth]{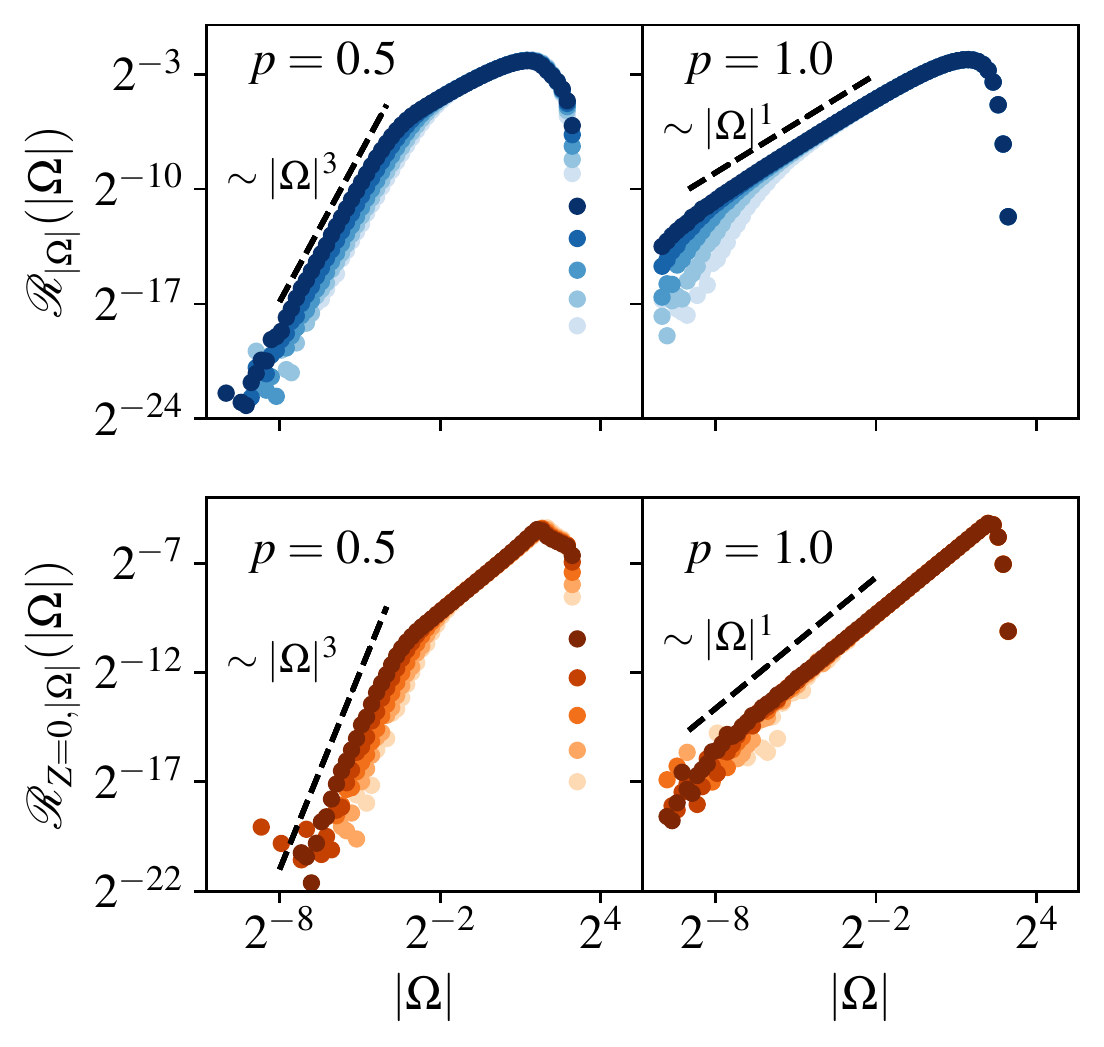}
\caption{Spectral correlations, $\mathscr{R}_{|\Omega|}$ (top row) defined in Eq.~\ref{eq:R},  and $\mathscr{R}_{Z=0,|\Omega|}$ (bottom row) defined in Eq.~\ref{eq:RZ0-modOmega}. The left and right columns represent the delocalised and localised phases respectively. In both the quantities, we see a distinct change from $|\Omega|^3$ (delocalised) to $|\Omega|$ (localised).}
\label{fig:RmodOmega}
\end{figure}

The normalised two-point spectral correlation is defined as 
\eq{
\mathscr{R}(\Omega) = \frac{1}{N(N-1)}\Braket{\sum_{\alpha\neq \beta}\delta(z_\alpha-z_\beta-\Omega)}\,.
\label{eq:R-Omega-complex}
}
As we did with the eigenvector correlations, we define the radial variation of the spectral correlations in the complex $\Omega$ plane by defining $\mathscr{R}_{|\Omega|}$ as
\eq{
\mathscr{R}_{|\Omega|} = |\Omega|\int_0^{2\pi} d\theta~ \mathscr{R}(\Omega)\,,
\label{eq:R}
}
with $\Omega=|\Omega|e^{i\theta}$.
Additionally, one can also define
\eq{
\mathscr{R}_{Z=0}(\Omega) = \frac{1}{N_{Z=0}}\Braket{\sum_{\alpha\neq \beta}\delta(z_\alpha+z_\beta)\delta(z_\alpha-z_\beta-\Omega)}\,,
\label{eq:RZ0}
}
where $N_{Z=0}$ is the number of pairs of eigenvalues with zero mean,
and the corresponding 
\eq{
\mathscr{R}_{Z=0,|\Omega|} = |\Omega|\int_0^{2\pi} d\theta~ \mathscr{R}_{Z=0}(\Omega)\,.
\label{eq:RZ0-modOmega}
}
In Fig.~\ref{fig:RmodOmega} we show the numerical results for $\mathscr{R}_{|\Omega|}$ (top row) and $\mathscr{R}_{Z=0,|\Omega|}$ (bottom row), both in the delocalised (left column) and localised (right column) phases. The spectral correlations are starkly different between the delocalised and localised phases in the universal regime of $|\Omega|\ll 1$. In the former, we find
\eq{
    \mathscr{R}_{|\Omega|},R_{Z=0,|\Omega|}\sim |\Omega|^3\,,
}
whereas in the localised phase, 
\eq{
    \mathscr{R}_{|\Omega|},\mathscr{R}_{Z=0,|\Omega|}\sim |\Omega|\,.
}

The result in the delocalised phase is again, just as for eigenvector correlations, symptomatic of the GinUE universality class. From the seminal paper by Ginibre~\cite{ginibre1965statistical}, the marginal joint distribution of two eigenvalues is given by
\eq{
P_\mathrm{GinUE}(z_{\alpha},z_\beta)\sim (1-&e^{-N|z_\alpha-z_\beta|^2})\times\nonumber\\&\Theta(1-|z_\alpha|)\Theta(1-|z_\beta|)\,,
\label{eq:P2-ginue}
}
where we have assumed that the eigenvalues are uniformly distributed over the unit disk and we have neglected the normalisation factors since we are interested only in the scaling. Using Eq.~\ref{eq:P2-ginue} in Eq.~\ref{eq:R} and in Eq.~\ref{eq:RZ0}, the spectral correlations can be obtained as 
\eq{
\mathscr{R}(\Omega) \sim \int dz_\alpha\int dz_\beta P(z_\alpha,z_\beta)&\delta(z_\alpha-z_\beta-\Omega)\,,
}
\eq{
\mathscr{R}_{Z=0}(\Omega) \sim \int dz_\alpha\int dz_\beta P(z_\alpha,&z_\beta)[\delta(z_\alpha+z_\beta)\times\nonumber\\&\delta(z_\alpha-z_\beta-\Omega)]\,.
}
Evaluation of the integrals yields
\eq{
\mathscr{R}(\Omega)\sim (1-e^{-N|\Omega|^2})\left[\mathrm{cos}^{-1}\left(\frac{|\Omega|}{2}\right)-\dfrac{|\Omega|}{2}\sqrt{1-\frac{|\Omega|^2}{4}}\right]\,,
\label{eq:RGinUEres}
}
and 
\eq{
\mathscr{R}_{Z=0}(\Omega)\sim (1-e^{-4N|\Omega|^2})\Theta(2-|\Omega|)\,.
\label{eq:RZ0GinUEres}
}
In the limit of $|\Omega|\ll 1$, both Eq.~\ref{eq:RGinUEres} and Eq.~\ref{eq:RZ0GinUEres} yield,
\eq{
\mathscr{R}_{|\Omega|}(|\Omega|),\mathscr{R}_{Z=0,|\Omega|}(|\Omega|)\sim |\Omega|^3\,,
}
and this is reflected in the numerical results in Fig.~\ref{fig:RmodOmega} (left).

Deep in the localised phase, by contrast, the eigenvalues can be well approximated as uncorrelated random numbers,
\eq{
P_\mathrm{loc}(z_{\alpha},z_\beta)\sim \Theta(1-|z_\alpha|)\Theta(1-|z_\beta|)\,,
\label{eq:P2-loc}
}
such that the spectral correlations take the form 
\eq{
\mathscr{R}(\Omega)\sim \left[2\mathrm{cos}^{-1}\left(\frac{|\Omega|}{2}\right)-|\Omega|\sqrt{1-\frac{|\Omega|^2}{4}}\right]\,,
}
and 
\eq{
\mathscr{R}_{Z=0}(\Omega)\sim \Theta(2-|\Omega|)\,.
}
They asymptotically yield, in the limit of $|\Omega|\ll 1$,
\eq{
\mathscr{R}_{|\Omega|}(|\Omega|),\mathscr{R}_{Z=0,|\Omega|}(|\Omega|)\sim |\Omega|\,,
}
which is consistent with the numerical results in Fig.~\ref{fig:RmodOmega} (right).

\end{document}